\documentclass[12pt,epsf,amstex]{article}
\usepackage [dvips]{graphicx}
\usepackage{amsmath}
\usepackage{amssymb}
\usepackage{epsfig}
\usepackage{verbatim}

\usepackage{color}

\addtocounter{secnumdepth}{1}
\font\capfont=cmbx12 at 50 pt %
\newbox\capbox \newcount\capl \def\a{A}
\def\docappar{\medbreak\noindent\setbox\capbox\hbox{%
\capfont\a\hskip0.15em}\hangindent=\wd\capbox%
\capl=\ht\capbox\divide\capl by\baselineskip\advance\capl by1%
\hangafter=-\capl%
\hbox{\vbox to8pt{\hbox to0pt{\hss\box\capbox}\vss}}}
\def\cappar{\afterassignment\docappar\noexpand\let\a }

\begin{document}

\newcommand{\ee}{{\rm e}}
\newcommand{\dd}{{\rm d}}
\newcommand{\p}{\partial}
\newcommand{\calT}{{\cal T}}

\newcommand{\Jf }{J^{\rm free}}
\newcommand{\Jj }{J^{\rm J}}
\newcommand{\Jr }{J^{\rm R}}
\newcommand{\Jff}{J^{\rm FF}}
\newcommand{\Jfj}{J^{\rm FJ}}
\newcommand{\Jjf}{J^{\rm JF}}
\newcommand{\Jjj}{J^{\rm JJ}}
\newcommand{\JM}{J^{M}}

\newcommand{\rhof }{\rho^{\rm free}}
\newcommand{\rhoj }{\rho^{\rm jam}}
\newcommand{\rhoff}{\rho^{\rm FF}}
\newcommand{\rhofj}{\rho^{\rm FJ}}
\newcommand{\rhojf}{\rho^{\rm JF}}
\newcommand{\rhojj}{\rho^{\rm JJ}}

\newcommand{\LR}{L^M_{{\rm R},m}}
\newcommand{\NR}{N^M_{{\rm R},m}}
\newcommand{\vR}{v^M_{{\rm R},m}}

\newcommand{\vf}{v^{\rm free}}
\newcommand{\fO}{f_E}
\newcommand{\fp}{f_P}
\newcommand{\fb}{f_B}
\newcommand{\fa}{f_A}

\newcommand{\Imx}{I^{mx}}
\newcommand{\Imy}{I^{my}}
\newcommand{\Imdelta}{I^{m\delta}}
\newcommand{\ttau}{\tilde{\tau}}
\newcommand{\tk}{\tilde{k}}
\newcommand{\sfR}  {{\sf{R}}}
\newcommand{\sfRm} {{\sf{R}}_m}
\newcommand{\sfRM} {{\sf{R}}^{M}}
\newcommand{\sfRMm}{{\sf{R}}^{M}_m}
\newcommand{\sfvMm}{{\sf{v}}^{M}_m}
\newcommand{\sfLMm}{{\sf{L}}^{M}_m}
\newcommand{\sfNMm}{{\sf{N}}^{M}_m}

\newcommand{\kdeltam}{k^{m\delta}}
\newcommand{\Tdeltam}{T^{m\delta}}
\newcommand{\taudelta}{\tau^{\delta}}
\newcommand{\taudeltam}{\tau^{m\delta}}

\newcommand{\alphac}{\alpha^{\rm c}}
\newcommand{\alphaM}{\alpha^{M}}
\newcommand{\alphaMm}{{\alpha^{M}_m}}
\newcommand{\alphaten}{\alpha^{10}}
\newcommand{\betaM}{\beta^{M}}
\newcommand{\betaMm}{{\beta^{M}_m}}

\newcommand{\la}{\langle}
\newcommand{\ra}{\rangle}
\newcommand{\beq}{\begin{equation}}
\newcommand{\eeq}{\end{equation}}
\newcommand{\bea}{\begin{eqnarray}}
\newcommand{\eea}{\end{eqnarray}}
\def\lsim{\:\raisebox{-0.5ex}{$\stackrel{\textstyle<}{\sim}$}\:}
\def\gsim{\:\raisebox{-0.5ex}{$\stackrel{\textstyle>}{\sim}$}\:}

\numberwithin{equation}{section}

\thispagestyle{empty}
\title{\Large  
{\bf A multi-lane TASEP model}\\[2mm]
{\bf for crossing pedestrian traffic flows}\\
\phantom{XXX}
}

\author{{H.J. Hilhorst and C. Appert-Rolland}\\[5mm]
{\small Laboratoire de Physique Th\'eorique, b\^atiment 210}\\
{\small Universit\'e Paris-Sud and CNRS,
91405 Orsay Cedex, France}\\}

\maketitle
\begin{small}
\begin{abstract}
\noindent
A one-way {\it street\,} of width $M$ is modeled as a set of
$M$ parallel one-dimensional TASEPs.
The intersection of two perpendicular streets
is a square lattice of size $M\times M$.
We consider hard core particles entering each street  
with an injection probability $\alpha$.
On the intersection square the hard core exclusion creates a
many-body problem of strongly interacting TASEPs
and we study the collective dynamics that arises.
We construct an efficient algorithm 
that allows for the simulation of streets of infinite length, 
which have sharply defined critical jamming points.
The algorithm employs the `frozen shuffle update',
in which the randomly arriving particles
have fully deterministic bulk dynamics.
High precision simulations for street widths up to $M=24$
show that when $\alpha$ increases, there occur jamming transitions 
at a sequence of $M$ 
critical values $\alphaM_M<\alphaM_{M-1}<\ldots<\alphaM_1$.
As $M$ grows, the principal transition point $\alphaM_M$ decreases roughly
as $\sim (\log M)^{-1}$ in the range of $M$ values studied.
We show that a suitable order parameter is provided by
a reflection coefficient associated with the particle
current in each TASEP.
\\

\end{abstract}
\end{small}
\vspace{12mm}

\noindent LPT Orsay 12/34
\newpage


\section{Introduction} 
\label{sect_introduction}

\cappar Pedestrian motion in dense environments
is of both theoretical and practical interest.
Instances of applications are shopping streets, waiting lines, 
crowds that enter or leave a confined space, and so on.
Under such circumstances 
simplified models may help understand the behavior of individuals 
as well as the collective behavior that results from it
\cite{schadschneider2008b,helbing2001b,bellomo_d2011,vicsek_z2012}.
A particular class of such models is based on cellular automata
\cite{burstedde2001b,schadschneider2002a,kirchner_n_s2003,klupfel2007}.
For modeling unidirectional {\it one-dimensional\,} traffic,
whether it be of particles, vehicles, or pedestrians \cite{chowdhury_s_s2000},
one popular tool is the
Totally Asymmetric Simple Exclusion Process (TASEP) \cite{derrida1998c}.
This stochastic process 
belongs to the larger class of models of random walkers
with hard core interactions and various different TASEP versions
have been studied by physicists
and mathematicians alike since several decades.
Specifically, a TASEP is a system of hard core particles that advance 
along a linear lattice in a single direction. 
The TASEP may be used as a building block
for traffic flow models in more complicated geometries.
There is a rich literature on junctions and bifurcations
\cite{brankov_p_b04,pronina_k05,huang05,huang06}, 
as well as on intersections 
\cite{nagel_s92,fouladvand_s_s04b,foulaadvand_na07,du_etal10,%
foulaadvand_b11,ARCH11c}
of TASEPs.
In those approaches the interaction between the different
TASEPs is `weak' in the sense that they are coupled
only on a set of generally well-separated sites.
Strongly interacting TASEPs were considered, in particular, in 
the so-called BHL model proposed by Biham {\it et al.}
\cite{Bihametal92}, 
which in turn has given rise to an offspring of variants
on some of which we will comment.
\vspace{3mm}

Although pedestrian traffic flow is our basic motivation, 
we present below a model which, because of its simplicity,
has an intrinsic interest that extends beyond this particular
application. From a wider perspective it is
an example of a driven nonequilibrium system, and,
contrary to many other such systems
that have been studied in the literature and that
have stochastic dynamics \cite{schmittmann_z95,schmittmann_z98}, 
this one is deterministic.
\vspace{3mm}

\begin{figure}
\begin{center}
\scalebox{.55}
{\includegraphics{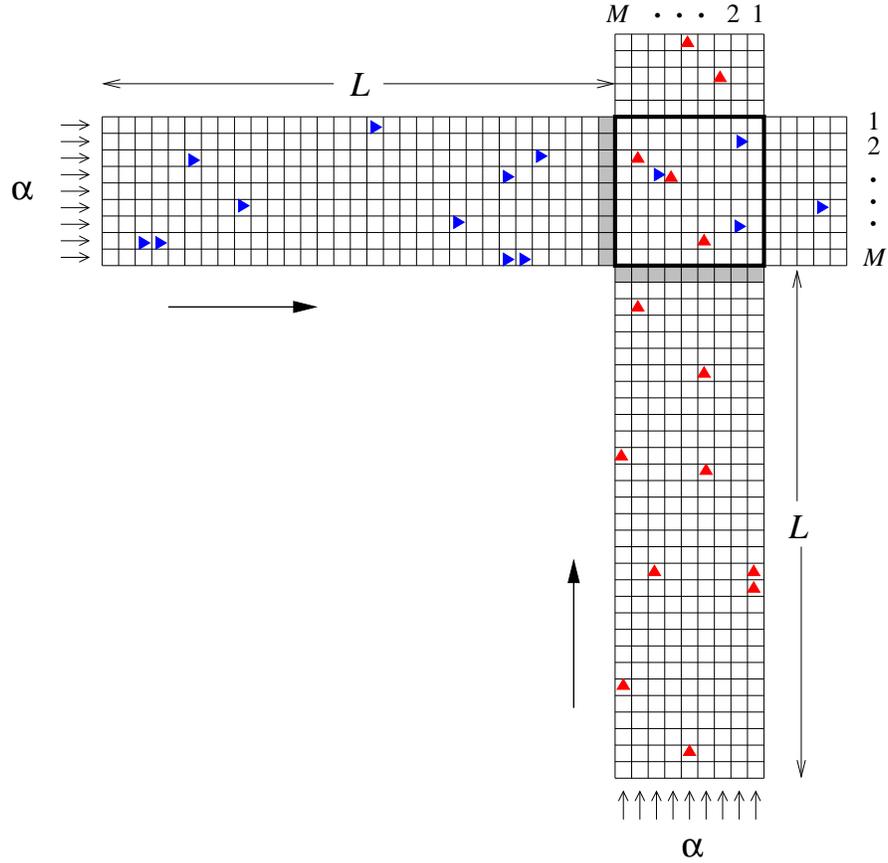}}
\end{center}
\caption{\small Intersection of two one-way streets of width $M$. 
The $x$ particles, moving to right, and the $y$ particles, moving upward, 
are represented by blue and red triangles, respectively.
The control parameter $\alpha$ is the injection probability 
per time step for each lane. 
Small arrows point towards the `injection' sites.
Lanes are numbered $m=1,2,\ldots,M$.
The region bordered by the heavy solid line is the `intersection
square'. The `entrance sites' defined in section \ref{sect_reduced} 
are shaded.}
\label{fig_intersectinglanes}
\end{figure}

We define a {\em street of width $M$}
as a set of $M$ parallel linear lattices to be
called {\em lanes}, each of which carries a TASEP.  
We wish to study what happens when two
such streets, both thought of as being of infinite length, 
intersect perpendicularly. Figure \ref{fig_intersectinglanes}
shows the geometry for the finite length case.
Particles in the horizontal street 
are injected from the left onto empty sites
with a probability $\alpha$ per time step.
Particles in the vertical street  
are injected from below with the same probability.
The resulting average incoming current in a lane 
will be denoted by $\Jf(\alpha)$.

The intersection area of the two streets is a
finite $M\times M$ square lattice.
On this square we have a problem of {\em strongly\,} interacting TASEPs,
whose study is the purpose of this work.
Lanes are numbered by an index $m=1,2,\ldots,M$ 
from the outer ones inward.
Unless symmetry were broken spontaneously, 
the $m$th horizontal and $m$th vertical
lane are statistically identical. 
One basic question here, as in any other traffic flow model, is to determine 
the outgoing currents $J^M_m(\alpha)$ as a function of $\alpha$
when the system is in a stationary state. 
Contrary to the incoming current, the
outflow must be expected to be $m$ and $M$ dependent. 
A lane is said to be in a {\em free flow phase\,} when its 
outgoing current is equal to the incoming current,
that is, when $J^M_m(\alpha)=\Jf(\alpha)$. 
At sufficiently low $\alpha$ free flow is to be anticipated 
for all $m$;  however, it is inevitable that
when $\alpha$ increases above a certain threshold,
part or all of the lanes undergo a jamming transition.
\vspace{3mm}

The update scheme is an essential part of the definition of any TASEP. 
We choose to employ here the frozen shuffle update
\cite{ARCH11a,ARCH11b,ARCH11c}, whose characteristic is that
in each time step the particles are updated according to a
sequence fixed once and for all. 
Particles entering the system are inserted in this
sequence and particles leaving are deleted from it.
This update, originally 
proposed as an alternative to parallel update%
\footnote{Frozen shuffle update may also be seen as a variant of 
{\it random\,} shuffle update \cite{wolki_s_s06,smith_w07}; 
in the latter
a new random particle order is drawn before each time step.},
has the advantage 
of eliminating the algorithmic conflicts that arise when in the same time
step two particles have the same target site.
It conserves the advantage of the parallel update in that it bounds 
the fluctuations that random sequential update would cause.
In our present implementation of the frozen update scheme all allowed moves
are carried out with probability one and hence
the bulk dynamics is that of a cellular automaton: it is fully deterministic.
\vspace{3mm}

Under frozen shuffle update,
once a particle has left the intersection square it continues
unimpeded at unit speed, and hence the street segments
beyond the intersection square need not be considered.
The street segments leading up to this square, however,
play an essential role,
since they are the place where waiting lines may develop, whether temporarily
or permanently. Such waiting lines are an intrinsic part of the 
traffic flow problem and we wish to fully account for them.
In any simulation the ingoing street segments 
are necessarily 
of finite length $L$, as shown in figure \ref{fig_intersectinglanes}.
A key observation is, however, that
sharply defined jamming
transition points on the $\alpha$ axis can exist
only in the limit $L\to\infty$. 
Indeed, as previous work \cite{ARCH11c} has shown, 
there is a sharp transition point 
for crossing streets even of width $M=1$, as long as their length 
$L$ is infinite. 
This being so, the present work does the following.

First, by combining theoretical and algorithmic arguments,
we show that it is possible to integrate out
the degrees of freedom in the half-way infinite
street segments leading up to the intersection square.
There then results an
efficient algorithm that simulates the particle motion on the 
{\it finite\,} $M\times M$ square, subject to boundaries 
on its left and lower edge that represent 
the incoming street segments of infinite length, $L=\infty$.
Finite size effects have thus been eliminated.
The appropriate boundary conditions are derived and formulated in terms of
{\it memory variables\,} 
to be defined in section \ref{sect_reduced}.

Secondly, we present the results of the simulation
of this interacting street model
as a function of its two parameters $\alpha$ and $M$.
We obtain the outgoing currents $J^M_m(\alpha)$ and find that
when $\alpha$ increases from $0$ to $1$, there appear $M$ critical values 
\beq
\alphaM_{M}<\alphaM_{M-1}<\ldots<\alphaM_{1}
\label{criticalalphas}
\eeq
at which, successively, the lanes 
with indices $m=M,M-1,\ldots,1$ become jammed. 
We will refer to $\alphaM_{M}$ 
as the {\em principal critical point\,} of the size $M$ intersection.
For $\alpha$ beyond the critical value $\alphaM_{m}$,
the outflow of particles in the $m$th horizontal and vertical lane
cannot keep up with the incoming current and in those lanes
ever growing waiting lines develop. 

We relate the incoming and outgoing current in lane $m$ by
\beq
\JM_m = (1-\sfRM_m)\Jf. 
\label{relJmJf}
\eeq
The product $\sfRM_m\Jf$ in (\ref{relJmJf}) may be interpreted 
as the reflected current%
\footnote{This even though all particles move
only in a single direction; see section \ref{sect_domainwall}.}
in the $m$th lane and we therefore call $\sfRM_m$
its {\it reflection coefficient}. 
We will show that there exists a simple theoretical
relation between this  coefficient 
and the memory variables occurring in the boundary condition. 
In our Monte Carlo work we deduce the critical point values $\alphaMm$  
from the $\alpha$ dependence of $\sfRMm$ through the criterion
\bea
\sfRM_m(\alpha) = 0, &&\qquad \alpha \leq \alphaM_{m}\,,  \nonumber\\[2mm]
\sfRM_m(\alpha) > 0, &&\qquad \alpha    > \alphaM_{m}\,,  
\qquad m=1,2,\ldots,M. 
\label{signsfR}
\eea
For $\alphaM_{m}<\alpha<1$ the coefficient $\sfRM_m(\alpha)$
increases monotonously and we may appropriately
consider it as an order parameter. 
The main results of our simulations are the determination of the
critical points $\alphaM_m$ and of the curves $\sfRM_m(\alpha)$.
\vspace{3mm}

This paper is set up as follows. 
In section \ref{sect_full} we define the rules of motion
for what we will call the `full' algorithm, that is, the one executed
on the lattice of figure \ref{fig_intersectinglanes}, which has a finite $L$.
In section \ref{sect_reduced} the memory variables are introduced;
we then show how for $L=\infty$
the full algorithm gives rise to a `reduced'  algorithm,
restricted to the $M\times M$ interaction square, and
with boundary conditions formulated in terms of these new variables.
In section \ref{sect_averages} we establish the theoretical expressions,
valid for each lane separately, that relate
the memory variable to the transmitted current and the speed
of propagation of the reflected current.
In section \ref{sect_simulations} we present and discuss simulation
results for lattices up to $M=24$. 
In section \ref{sect_conclusion} 
we make several final remarks and conclude.


\section{Intersecting street model}
\label{sect_full}

We consider the lattice of figure \ref{fig_intersectinglanes}
showing a `horizontal' and a `vertical' 
one-way street labeled 'x' and 'y', respectively,
and each composed of $M$ parallel lanes.
A lane is identified by an index pair $(m,\delta)$ with
$m=1,2,\ldots,M$ and $\delta=\mbox{`$x$'}$ or $\mbox{`$y$'}$.
The length $L$ of the street segments
leading up to the intersection  
is a large finite number%
\footnote{In section \ref{sect_reduced} we will set $L=\infty$.}.
The two streets intersect according to a square of $M\times M$ sites. 
Two types of hard-core particles move on the lattice, $x$ particles 
arriving from the left along the horizontal street, and $y$ particles
arriving from below along the vertical street.
New particles may be injected onto
the sites of the leftmost column of the horizontal street 
and of the bottom row of the vertical street; 
we will refer to these as the {\it injection\,} sites.
We let $a>0$ denote the {\it injection rate\,} of particles onto an empty
injection site. The ensuing {\it injection probability}
$\alpha=1-\ee^{-a}$ is the probability that an injection site,
when empty at a certain instant of time $t$,
will get occupied during the unit time interval $(t,t+1)$ that follows.
We will employ $\alpha$ rather than $a$ as the control parameter.


\subsection{Full algorithm}
\label{sect_fullalgorithm}

The rules for the particle motion on this lattice are as follows. 
The algorithm is a succession of time steps labeled by an
integer time step index $s=1,2,3,\ldots$. 
The $s$th time step takes the system from time $t=s-1$ to $t=s$.
It executes all events (particle moves, injections and
deletions) that occur during the time interval $s-1 < t < s$.

With the frozen shuffle update  \cite{ARCH11a,ARCH11b,ARCH11c} employed here,
either at the initial time $t=0$, or else when it enters the system,
the $i$th particle in lane $(m,\delta)$ is assigned, 
in a way to be described below, 
a random {\it phase\,} $\taudeltam_i\in(0,1)$;
it keeps this phase as a fixed attribute.
The phases define in which order the particles are updated at each
time step.
For practical reasons the algorithm
decouples each time step update
into a `bulk' step and an `injection' step, both to be described now.


\subsubsection {Bulk step}
\label{sect_bulkstep}

The $s$th bulk time step consists of a sweep through 
the `update sequence',
that is a list, common to all lanes, 
of all particles
present in the system at the beginning of that time step,
ordered according to increasing phases.
Thus, during the bulk step, each particle present in the
system is visited 
exactly once.
This bulk step parallels the evolution of the
continuous time variable $t$, the turn for particle $i$ to be updated
being identified with the instant of time $t=s-1+\taudeltam_i$.
\vspace{2mm}

{\it General case.\,\,}
The case of a general particle $i$ in a lane $(m,\delta)$ is simple.
When at time  $t=s-1+\taudeltam_i$ its turn to be updated has come,
then if at that instant of time its target site is occupied,
it does not move; if its target site is empty, it moves. 

{\it First special case.\,\,}
A particle $i$ occupying the last site of its lane 
(an `exit site') has no target site. 
When its turn to be updated has come, it is deleted from the system; 
in particular, its phase is deleted from the update sequence.
We will comment further on this boundary condition in section 
\ref{sect_exit}.

{\it Second special case.\,\,}
When a particle $j$ moves off the injection site of
a lane $(m,\delta)$,
say at time $s-1+\taudeltam_{j}$\,, the algorithm needs to do more work.
In that case,
in order to prepare for the injection of the next particle, $j+1$, 
onto that injection site, a random time interval $\Tdeltam_{j+1}$ is drawn 
from the exponential distribution
\beq
P(T) = a\,\ee^{-aT}, \qquad T > 0,
\label{defPT}
\eeq
and $j+1$ is scheduled to be injected at time $t_{j+1}$ given by%
\footnote{When no confusion can arise we will suppress the superscript
`$m\delta$' on $\taudeltam_j$. Similarly, we will not append this
superscript on quantities like $t_{j+1}$ and $k_{j+1}$ introduced in this 
subsection.}
\beq
t_{j+1} = s-1+\tau_{j}+\Tdeltam_{j+1}\,.
\label{xtjp1}
\eeq
This implies, first, 
that this new particle has a phase $\tau_{j+1}$ equal to
the fractional part of its time of injection, that is,
\beq
\tau_{j+1} = (\tau_{j}+\Tdeltam_{j+1})\,\mbox{mod}\,1;
\label{rectaui}
\eeq
and, secondly, that its injection will occur during time step
$s_{j+1} = \lfloor t_{j+1}\rfloor +1 
= s +\lfloor \tau_j + \Tdeltam_{j+1}\rfloor$,
where $\lfloor x\rfloor$ denotes the largest integer contained in $x$.
Particle $j+1$, its phase $\tau_{j+1}$, 
its time step of injection $s_{j+1}$,
and its lane index $(m,\delta)$, 
are placed on a waiting list to which the algorithm will return 
during the injection step. 
This completes the discussion of the second special case.
\vspace{3mm}

{\it Remark\,} (i).\,\, 
We have $s_{j+1}\geq s$, that is, the injection
may occur either during the same time step $s$ in which the injection 
site was emptied, or during any later time step.

{\it Remark\,} (ii).\,\,
The injection rule implements independent and uniformly
distributed arrival times subject to interparticle hard core exclusion
\cite{ARCH11b,ARCH11c}.


\subsubsection {Injection step}
\label{sect_injectdelete}

The waiting lists of each of the $2M$ lanes $(m,\delta)$ 
are consulted independently. If the particle, say $k$, on that list
has a time step of injection $s_k$ equal to the current time step $s$,
then the algorithm implements the injection; the phase $\taudeltam_{k}$
is inserted in the update sequence; and the injection is considered to
have occurred physically at time $s-1+\taudeltam_{k}$. 
In all following time steps,
and until particle $k$ has left the system, its position 
will be updated as part of the bulk step.
\vspace{3mm}

{\it Remark.\,\,} 
In the absence of any obstacles to the particle flow
the above injection procedure brings the incoming lane segment into a
`free flow configuration', that is, one in which each
particle advances by one lattice unit each time step%
\footnote{A `free flow configuration' is a microstate.
Note that the macroscopic `free flow phase' was
defined in section \ref{sect_introduction} by the equality
of inflow and outflow.}.
This configuration has an average particle density
$\rhof(\alpha)$ and an average current $\Jf(\alpha)=v\rhof(\alpha)$ given by 
\cite{ARCH11b,ARCH11c}
\beq
\rhof(\alpha)=\frac{a}{1+a}\,, \qquad \Jf(\alpha)=\frac{a}{1+a}\,,
\label{xrhoJF}
\eeq
where we used that $v=1$.


\subsubsection{Initial state}
\label{sect_initial}

At time $t=0$ we initialize the system in a random free flow configuration
in which the flow has just reached the entrance sites 
(shown shaded in figure \ref{fig_intersectinglanes})
of the intersection square, but has not entered the square itself yet.
Such a configuration is conveniently generated as follows.

(1) Set time equal to $t=-L+1$.

(2) On each of the injection sites
of the $2M$ lanes independently,
deposit a particle with the probability $\rhof(\alpha)$ given by
(\ref{xrhoJF}) and leave it unoccupied
with the complementary probability $1-\rhof(\alpha)$.
If a particle is deposited,
assign to it a phase $\taudeltam_0$ drawn randomly and uniformly from $(0,1)$.
If no particle is deposited, draw a $\Tdeltam_1$ from (\ref{defPT}) and
prepare to occupy the injection site at a time $t=-L+1+\Tdeltam_1$  
(that is, during some time step $s\geq -L+2$)
by a particle of phase $\taudeltam_1=\Tdeltam_1\,\mbox{mod}\,1$.

(3) Execute $L-1$ time steps to take the system from time $t=-L+1$
to $t=0$. 
During these time steps the $2M$ lanes do not interact and no blocking occurs.
As a result, each particle initially deposited on an injection
site will at $t=0$ occupy the corresponding
entrance site, 
and the lane segments leading up to the intersection
square will carry a free flow configuration.

{\it Remark.\,\,}
Both the random initial state and
the time evolution of the system are fully determined by 
the sequences of time intervals
\beq
\Tdeltam_{1}, \Tdeltam_{2}, \Tdeltam_{3}, \ldots, \qquad 
 m=1,2,\ldots,M, \qquad \delta=\mbox{`$x$' or `$y$'}, 
\label{seqTdeltam}
\eeq
which, in particular, determine the $\taudeltam_{j}$ for $j=1,2,3\ldots$
through (\ref{rectaui}).


\subsection{The exit boundary}
\label{sect_exit}

The lane segments beyond the intersection square are initially empty. 
The first particle to enter onto the first site of such a
lane segment will from that moment on proceed unblocked at speed $v=1$.
Each next particle will enter the same segment
only after its predecessor has left the first site and from then on
similarly proceeds at speed $v=1$.
Hence each outgoing lane segment carries a free flow configuration
and does not exert any feedback on the intersection square.
The boundary condition that we applied in section \ref{sect_bulkstep} --
namely to eliminate a particle which when its update time has come
no longer has a target site -- 
therefore correctly represents an outgoing free flow heading to infinity.
We will henceforth apply this boundary condition
at the immediate upper and right border of the intersection square.


\section{The intersection square}
\label{sect_reduced}

\begin{figure}
\begin{center}
\scalebox{.40}
{\includegraphics{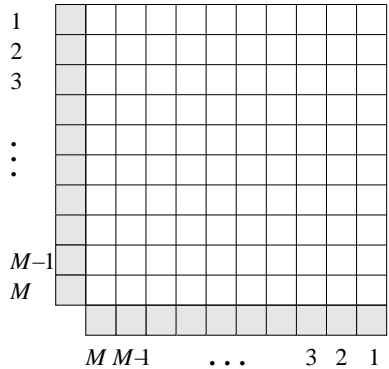}}
\end{center}
\caption{\small The $M\times M$ intersection square on which the
`reduced' algorithm is defined.
An extra row and column of `entrance sites' (shaded) 
along the left and lower border
serve to take care of the boundary conditions.}
\label{fig_intersectionsquare}
\end{figure}

We construct in this section an algorithm 
for simulating the crossing street model 
which is an exact consequence of 
the above `full' algorithm but which is much more efficient.
This `reduced' algorithm
retains only the particle motion on the intersection square.
As we have seen, the particles exiting this square cease to play a role.
We will show below that it is possible --
in the limit $L\to\infty$ (see figure \ref{fig_intersectinglanes}),
which is precisely the one of interest --
also to eliminate the incoming flows from the simulation.
This is achieved with the aid of appropriate
boundary conditions defined on the entrance sites
shown shaded in figure \ref{fig_intersectinglanes}.
Hence the system geometry becomes that of figure \ref{fig_intersectionsquare}. 
For $L=\infty$ all finite size effects are eliminated and the system
has sharply defined critical points. Any remaining uncertainty on the
location of these points will be due only to the finite duration of
the simulation.


\subsection{Memory variables $\Imdelta(s)$}
\label{sect_delayvariables}

All statistical properties of the free flow 
injected into a lane $(m,\delta)$ are known.
If this flow did not encounter any obstacles, then 
it would arrive unmodified at the entrance site $L$ time steps later,
and the boundary condition at that site would be known in a simple way.
However, blocking of particles in the intersection square
will typically create at the entrance site a waiting line 
of variable length which complicates the boundary conditions.
This waiting line may be either ever growing
or, instead, only fluctuating with a finite localization length,
two possibilities that correspond to the lane being in the `jammed flow'
and the `free flow' phase, respectively. 
The discussion of the present section applies to both cases.
 
The solution to the boundary condition problem
will be shown to consist in introducing
an auxiliary time dependent {\it memory variable\,} $\Imdelta(s)$. 
We will be able by means of this variable
to determine the arrival time $t=t_j$ of each particle $j$
at the entrance site {\it exactly as if,} starting at time $t=-\infty$,
it had made its way
through a street segment of length $L=\infty$, taking correctly into
account any time it may have spent blocked in the waiting line. 
\vspace{3mm}

We associate with each entrance site a `reference
particle': this will be the particle located on that site if there is one, 
and the first particle going to arrive on that site otherwise. 
Consider now the incoming segment of lane $(m,\delta)$.
During each time step every particle either
advances by one lattice unit or incurs a `unit time delay'.
We define the memory variable $\Imdelta(s)$ as 
the accumulated time delay that the reference particle in lane
$(m,\delta)$ has incurred in the course of its history
up to and including the $s$th time step.
It should be noted that $\Imdelta(s)$ is not associated with 
a fixed particle; when a particle leaves the entrance site,
the next one in the same lane takes over its role as the reference particle
and continues to carry the variable $\Imdelta(s)$,
the value of which is in general affected by this takeover process.
The $\Imdelta(s)$ are nonnegative integers; 
for the initial state described in subsection \ref{sect_initial} 
we have $\Imdelta(0)=0$. 
\vspace{2mm}


In order to determine the time evolution of the $\Imdelta(s)$
we examine in detail how
the $s$th time step relates $\Imdelta(s)$ to $\Imdelta(s-1)$.
In any time step 
three different events may occur
at the entrance site of lane $(m,\delta)$; we
denote them by the symbols `B' (for `blocked'), 
`A' (for `advancing'), and `E' (for `empty').

{\it Event\,} B.\,\,
The reference particle occupies the entrance site but
is blocked and does not move.
In that case its delay $\Imdelta(s)$ is augmented by one unit,
\begin{subequations}\label{dynImdelta}
\beq\label{dynImdelta1}
\Imdelta(s)=\Imdelta(s-1)+1.
\eeq 

{\it Event\,} A.\,\,
The reference particle, say $j$, occupies the entrance site and 
during the $s$th time step advances into the intersection square.
The next particle, $j+1$, becomes the reference particle
and $\Imdelta(s)$, now redefined
as the accumulated time delay incurred by $j+1$, must be recalculated. 
If $j$ would never have been blocked, its successor $j+1$
would have performed the same jumps as $j$ but
with a time delay $\Tdeltam_{j+1}+1$.
Here $\Tdeltam_{j+1}$ is
the difference between the departure time of $j$ and the arrival time of
$j+1$ on an arbitrary site, and the extra $+1$ is
the sojourn time of an unblocked particle on that site.

The extra headway $\Tdeltam_{j+1}$ allows $j$ to be blocked a total
number of $\lfloor\Tdeltam_{j+1}\rfloor$ times without blocking $j+1$.
Hence if the time delay $\Imdelta(s-1)$ of $j$ is small enough to satisfy
$\Imdelta(s-1) \leq \lfloor\Tdeltam_{j+1}\rfloor$, then $\Imdelta(s)=0$. 
Each supplementary blocking of $j$ leads to a blocking, and 
hence to a unit time delay, of $j+1$, and so if
$\Imdelta(s-1) \geq \lfloor\Tdeltam_{j+1}+1 \rfloor$, then
$\Imdelta(s) = \Imdelta(s-1) - \lfloor\Tdeltam_{j+1}\rfloor$.
This may be combined into
\beq\label{dynImdelta2}
\Imdelta(s) = \max\big( \Imdelta(s-1)-\lfloor \Tdeltam_{j+1}\rfloor,\,0 \big).
\eeq
We note that in event A 
particle $j+1$ may or may not arrive on the entrance site
during the time step $s$ under consideration.

{\it Event\,} E.\,\,
The reference particle does not occupy the
entrance site at the beginning of the $s$th time step.
Being free to move, it comes one lattice site closer
to the entrance site during that time step. 
Therefore its incurred time delay remains unchanged, that is,
\beq
\label{dynImdelta3} 
\Imdelta(s)=\Imdelta(s-1).
\eeq
\end{subequations}
Again, in this event the reference particle may or may not arrive on the
entrance site during time step $s$.
\vspace{2mm}

Equations (\ref{dynImdelta1})-(\ref{dynImdelta3})
govern the time evolution of $\Imdelta(s)$.
We are led to the important conclusion that
they involve exclusively the local motion of a single reference 
particle on or near the entrance site of lane $(m,\delta)$; 
there is therefore no need for simulating the half-way infinite lane segment
leading up to that site.
Allowing the integer $\Imdelta$, if needed, to increase without bound,
as we will do in the simulation, amounts to setting effectively
$L=\infty$. 
The reduced algorithm of the next subsection is based on 
these considerations.


\subsection{Reduced algorithm}
\label{sect_reducedalgorithm}

The reduced algorithm is an exact consequence of 
the full algorithm of section \ref{sect_initial} when the limit
$L\to\infty$ is taken. This limit, almost paradoxically, simplifies the
mathematics to the point that the reduced algorithm involves
only the particle positions in the $M\times M$ intersection
square and those in the row and column of entrance sites. 
There appear memory variables
$\Imdelta$ associated with these entrance sites.
As the original injection sites have moved to minus infinity, 
particle injection now takes place {\it de facto\,} on the entrance sites.
We state below only the points of difference between
the reduced algorithm and the full one described 
in section \ref{sect_initial}.


\subsubsection{Bulk step}
\label{sect_bulkstepB}

At the beginning of the $s$th time step, which covers the time
interval $s-1\leq t < s$, the memory variables
$\Imdelta(s-1)$ are known.
The bulk step consists again of a sweep through all particles, ordered
according to increasing phases, but now all located either on
the $M\times M$ square or on one of the entrance sites.
Particles move as in the full algorithm of subsection
\ref{sect_bulkstep}; the `general case' and the `first special case' 
of that section are the same here.

{\it Special case.\,\,} The only special case to be discussed
concerns the entrance sites. If during the $s$th time step
the sweep encounters
a particle $j$ that is blocked on an entrance site $(m,\delta)$, then
$\Imdelta(s)$ is calculated from $\Imdelta(s-1)$ according to 
(\ref{dynImdelta1}).

If during the $s$th time step, say at time $s-1+\tau_{j}$,
the sweep encounters a particle $j$ that
advances from the entrance site into the intersection square,
then a random $\Tdeltam_{j+1}$ is drawn as in the full algorithm and 
$\Imdelta(s)$ is determined according to (\ref{dynImdelta2}). 
The difference $\Imdelta(s-1)-\Imdelta(s)$ represents the reduction 
-- compared to the free flow situation -- of the number of time steps 
separating the departure of $j$ and the
arrival of $j+1$ on the entrance site.
Hence, recalling that $s-1+\tau_j$ is the time at which $j$ leaves 
the entrance site, we find that $j+1$ must be
injected onto that site at time $t_{j+1}$ given by
\bea
t_{j+1} &=& s-1 + \tau_j + \Tdeltam_{j+1} + \Imdelta(s) - \Imdelta(s-1)
\nonumber\\[2mm]
        &=& s-1 + \tau_j + \Tdeltam_{j+1} 
            - \min\big( \Imdelta(s-1),\lfloor\Tdeltam_{j+1}\rfloor \big),
\label{xsprime}
\eea
where to pass to the second line we used (\ref{dynImdelta2}).
Equation (\ref{xsprime}) may be compared to (\ref{xtjp1}).
The time step of injection is
$s_{j+1} = \lfloor t_{j+1}\rfloor +1  = 
s+\lfloor\tau_j+\Tdeltam_{j+1}\rfloor
- \min\big( \Imdelta(s-1),\lfloor\Tdeltam_{j+1}\rfloor \big)$. 
Particle $j+1$, its phase $\tau_{j+1}$, its time step of injection
$s_{j+1}$, and its lane index $(m,\delta)$
are placed on a list of particles waiting to be injected.

Finally, an entrance sites that is empty is not involved in the sweep.
For the corresponding lane $(m,\delta)$
the memory variable remains unchanged, that is,
$\Imdelta(s)=\Imdelta(s-1)$, in agreement with (\ref{dynImdelta3}).


\subsubsection{Injection step}
\label{sect_injectdeleteB}

Particle injection now takes place on the $2M$ entrance sites, to each
of which the injection procedure is applied independently.
If for a given entrance site $(m,\delta)$
the index $s$ of the current time step is equal to the
time step index of the next particle to be injected on that site,
say $k$, then the injection is carried out; 
the phase $\tau_{k}$
is inserted in the update sequence; and the injection is considered to
have occurred at time $s-1+\tau_{k}$. 


\subsubsection{Initial state}
\label{sect_initialB}

The initial state is generated as for the full model, 
section \ref{sect_initial}, but with $L$ replaced by $1$,
that is, the injection sites coincide with the entrance sites 
and step (3) of section \ref{sect_initial} is empty.
We set $\Imdelta(0)=0$ for all $(m,\delta)$.


\section{Jamming transitions}
\label{sect_averages}

Whereas the preceding sections 
have dealt with the microscopic algorithm,
the present section is of a theoretical nature. We 
derive certain relations that connect the averages directly
obtained in the simulation to other physically meaningful averages. 
As observed in the
remark that closes section \ref{sect_initial},
both the initial state and the dynamics are determined by the $2M$
sequences of independent interval variables $\Tdeltam_{j}$.
An average is therefore a mean value calculated or measured
with respect to the set $\{\Tdeltam_{j}\}$.
In a state with stationary currents%
\footnote{To which we will refer as the `stationary state', even though
  the memory variables, and concomitantly the waiting line, may or may
  not be stationary.}
an average over a sufficiently long
period of time must be expected to coincide with
this $\Tdeltam_{j}$ average.
Of primary interest will be the average currents $\JM_m(\alpha)$
for lane index $m=1,2,\ldots,M$ and as a function of the injection
probability $\alpha$.


\subsection{Current and reflection coefficient}
\label{sect_reflection}

For each lane index $m$ we must envisage two possibilities
whose actual occurrence is to be ascertained by the simulation.

(a) The incoming flow is weak enough so as to pass entirely,
which means that $\JM_m(\alpha)=\Jf(\alpha)$.
The lane is then in a free flow phase.

(b) The intersection square cannot handle the full
incoming flow, which means that $\JM_m(\alpha)<\Jf(\alpha)$.
The lane is then in a jammed flow phase.

We will write for either case
\beq
\JM_m = (1-\sfRM_m)\Jf,
\label{relcurrent}
\eeq
where $\sfRM_m(\alpha)$ is the {\it reflection coefficient\,} 
of the $m$th lane and has the properties (\ref{signsfR}).
The product $\sfRM_m\Jf$ may be interpreted 
as the reflected current in the $m$th lane, to be discussed in greater
detail in section \ref{sect_domainwall}.
Since $\Jf(\alpha)$, given by (\ref{xrhoJF}), 
is imposed and the $\JM_m(\alpha)$ may be measured
by counting the outgoing particles in each lane, substitution of these
two quantities in (\ref{relcurrent}) yields the $\sfRMm$.

There is, however, a direct way of determining the
$\sfRMm$ in the simulation. 
We will show below that 
in the long time limit the stochastic memory variable $\Imdelta(s)$ satisfies%
\footnote{Almost surely, in the mathematical sense.}
\beq
\lim_{t\to\infty} \frac{\Imdelta(t)}{t} = \sfRMm,
\qquad m=1,2,\ldots,M,
\label{defR}
\eeq
so that $\sfRMm$ is also the {\it average rate of growth\,} of 
this variable.


\subsection{Equivalence of (\ref{relcurrent}) and (\ref{defR})}
\label{sect_relImdeltaJ}

We will take (\ref{defR}) as the definition of
the $\sfRMm$ and show that (\ref{relcurrent}) follows.
In the case of free flow the waiting line fluctuates only within a
finite localization length, hence $\Imdelta(s)$ remains effectively
bounded whence $\sfRMm=0$, and (\ref{relcurrent}) is trivially true.
It suffices therefore to consider the case of a jammed phase, for which
$\Imdelta(t)$ is asymptotically linear in $t$. 
There are three steps to the proof.
\vspace{3mm}

{\it First step.\,\,}
In every time step $s$ one of the three equations (\ref{dynImdelta})
is applied to obtain $\Imdelta(s)$ from $\Imdelta(s-1)$.
In the stationary state, let $\fb$, $\fa$, and $\fO$ be the fractions%
\footnote{The notation leaves the
  dependence of these fractions on $m$ and $M$ implicit.}
of all time steps in which equations (\ref{dynImdelta1}),
(\ref{dynImdelta2}), and (\ref{dynImdelta3}), respectively, are applied.
When $\Imdelta(t)$ grows without bounds, the maximum
in equation (\ref{dynImdelta2}) is equal to
$\Imdelta(s-1)-\lfloor\Tdeltam_{j+1}\rfloor$.
It follows that for large times $t$ we have the asymptotic proportionality
\bea
\Imdelta(t) &\simeq& (0\times \fO + 1\times \fb - \kappa\times\fa)\,t,
\nonumber\\[2mm]
&=& (\fb-\kappa\fa)t,
\label{xImdelta}
\eea
in which $\kappa$ is the average of $\lfloor\Tdeltam_{j+1}\rfloor$ 
and is easily calculated as
\beq
\kappa = \int_0^\infty \!\dd T \lfloor T \rfloor P(T)
       = \frac{1-\alpha}{\alpha}
\label{xkappa}
\eeq
with $P(T)$ given by (\ref{defPT}).
Using (\ref{defR}), (\ref{xImdelta}), and (\ref{xkappa}) we deduce that
\beq
\sfRMm = \fb-\frac{1-\alpha}{\alpha}\,\fa\,, 
\label{xRMm}
\eeq
which completes the first step.
\vspace{3mm}


{\it Second step.\,\,} 
Using results of earlier work \cite{ARCH11b}
we now determine the event fractions $\fa$ and $\fb$ 
in terms of $\alpha$ and $\betaM_m$. 
We consider a particle on the entrance site.
Following \cite{ARCH11b} we let a 
parameter $1/\betaMm$ stand for the average number of time steps 
that such a particle has to wait  
before it can enter the intersection square%
\footnote{The inverse $\betaMm$ is the probability per time step that
a particle on the entrance site is allowed to enter the
intersection square. In reference \cite{ARCH11b} this parameter
(called $\beta$ there) was an
independent control parameter and entrance events of
successive particles were uncorrelated. In the present case $\betaM_m$
is a complicated function of
$\alpha$ and $M$; moreover, correlations must be expected between
entrance events.}.
A particle that leaves the entrance site 
during the $s$th time step
may or may not be replaced during the
same time step. It was shown in reference \cite{ARCH11b} that
if, in the jammed phase, this site is 
unoccupied at some integer time $s$, 
then it will certainly be reoccupied at time $s+1$.
Let $\nu$ be the average number of particles that cross the entrance
site%
\footnote{This sequence of particles is called a `platoon'.}
between two successive integer instants of time at which 
this site is unoccupied. This number $\nu$ is
determined exclusively by $\alpha$ and given by \cite{ARCH11b} 
\beq
\frac{1}{\nu} = 1 + \frac{1}{a} - \frac{1}{\alpha}\,.
\label{xnu}
\eeq
It follows that for each time step at which the entrance site is unoccupied
(event E above), 
there are on average $\nu/{\betaMm}$ time steps at which it is
occupied (events A and B above).
Out of the latter, there are $\nu$ time steps at which
the particle advances (event A), 
and a remaining $\left( 1/\betaMm -1 \right)\nu$ time steps at which
it stays blocked (event B). 
Hence when a lane is in the jammed state,
the event fractions $\fO$, $\fa$, and $\fb$ are given by
\beq
\fO= \frac{1}{\nu/\betaMm + 1}\,, \qquad
\fa = \frac{\nu}{\nu/\betaMm + 1}\,, \qquad
\fb = \frac{(1/\betaMm - 1)\nu}{\nu/\betaMm + 1}\,.
\label{deffractions}
\eeq
Substituting (\ref{deffractions}) in (\ref{xRMm}) we obtain
\beq
\sfRMm = \frac{\nu\betaMm}{\nu+\betaMm}
\left( \frac{1}{\betaMm}-\frac{1}{\alpha} \right).
\label{xsfR}
\eeq
Only for positive $\sfRMm$ 
is this result consistent with our initial supposition that lane $m$ is in the
jammed phase.  Therefore,
(\ref{xsfR}) shows that lane $m$ is jammed 
when $\alpha > \betaMm$. 
Remembering that $\betaMm$ is a function, although unknown, of 
$\alpha$ and $M$ we see that
the jamming point $\alpha=\alphaMm$ 
is the solution of $\betaMm(\alpha)=\alpha$. 
\vspace{3mm}


{\it Third step.\,\,} 
In the jammed phase
the particle density on any site of the waiting line,
and in particular on the entrance site, is given by
$\rhoj=\fa+\fb$ and the current entering
the intersection square is equal to $\betaMm\rhoj$.
Since in the stationary state, the current entering
the intersection square in lane $m$ is equal 
to the outgoing current $\JM_m$ in that lane, it follows with the aid of
(\ref{deffractions}) that
\beq
\rhoj(\alpha,\betaMm) = \frac{\nu}{\nu+\betaMm}, \qquad 
\JM_m(\alpha,\betaMm) = \frac{\nu\betaMm}{\nu+\betaMm}\,,
\label{xJbetam}
\eeq 
which is the jammed phase counterpart to (\ref{xrhoJF}).
Upon eliminating $\betaMm$ from 
(\ref{xsfR}) and (\ref{xJbetam}), solving for 
$\JM_m$ in terms of $\sfRMm$, and using
(\ref{xrhoJF}), we obtain (\ref{relcurrent}).
We have therefore established the interpretation of $\sfRMm$, 
initially defined by (\ref{defR}), 
as the reflection coefficient of the $m$th lane. 
This completes the equivalence proof of this subsection.
\vspace{3mm}

Equations (\ref{relcurrent}), (\ref{xsfR}), and (\ref{xJbetam}) 
show that out of the three quantities
$\sfRMm$, $\JM_m$, and $\betaM_m$, each one determines the two others.


\begin{figure}
\begin{center}
\scalebox{.45}
{\includegraphics{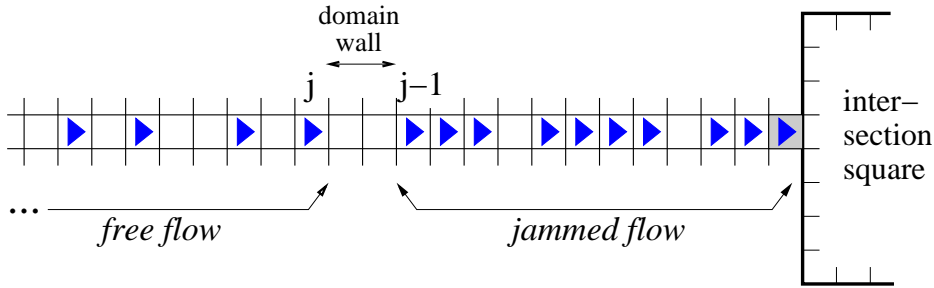}}
\end{center}
\caption{\small Lane containing a domain wall that separates a free flow 
from a jammed flow region. 
The particles in the latter constitute the waiting line.
The particle on the entrance site (shaded) is the reference particle
for the lane's memory variable.
}
\label{fig_domainwall}
\end{figure}

\subsection{Reflected current and domain wall motion}
\label{sect_domainwall}

We consider a single lane $m$
and suppose that $\alpha>\alphaM_m$ so that jamming will occur.
We now refer to figure \ref{fig_domainwall}. 
Going from left to right, the first particle in the waiting line 
is by convention the leftmost particle ever to have been blocked;
in the figure this is particle $j-1$%
\footnote{This 
cannot be concluded from the figure alone, since the property
of having never been blocked depends on the history of the
configuration.}.
We will say that a `domain wall' is located immediately to the left of 
this particle.
The waiting line constitutes a finite spatial domain of jammed flow,
separated by the domain wall
from a half-infinite region in the free flow phase.
The domain wall {\it motion\,} in a finite
one-dimensional system has been discussed in references \
\cite{ARCH11b,santen_a02}.
\vspace{3mm}

Our reduced algorithm no longer contains the
description of the domain walls.
We will show below, however, that if at time $s$ 
we know $\Imdelta(s)$ for some lane $(m,\delta)$, 
we can relate this quantity analytically to
the {\it average\,} position that the domain wall in that lane 
would have at that time. 
For the initial conditions of section \ref{sect_initialB} 
the domain wall is located at the entrance of the
intersection square and, for $\alpha>\alphaM_m$, starts 
at time $t=0$ propagating in the negative direction
at some average speed that we will call%
\footnote{Quantities related to this wall will be indicated by 
sans serif symbols.}
$\sfvMm$. 
If we assimilate the incoming particle flow to a wave, then
the domain wall is the moving front of the reflected wave. 

Consider now a time $T$ large enough for fluctuating variables to be 
approximated by their averages.
Let $\sfLMm$ be the linear size of the
jammed flow region, so that $\sfvMm=\sfLMm/T$.
To determine $\sfLMm$ we reason as follows.
The average number of particles $\sfNMm$ that up to time $T$
has been prevented to cross the interaction square is 
\beq
\sfNMm = T(\Jf-\JM_m).
\label{xNR}
\eeq
These particles are spread out along the waiting line
and therefore the density $\rhoj$ of this line is equal to $\rhof$ plus
the extra contribution $\sfNMm/\sfLMm$.
This leads to a continuity equation in the form
\beq
\rhoj(\alpha,\betaM_m) = \rhof(\alpha) + \frac{T}{\sfLMm}
\left[\Jf(\alpha)-\JM_m(\alpha,\betaM_m)\right],
\label{continuity}
\eeq
in which $\rhoj(\alpha,\betaM_m)$ and $\rhof(\alpha)$ are known from 
(\ref{xJbetam}) and (\ref{xrhoJF}), respectively.
Using that $\sfvMm=\sfLMm/T$ 
we deduce that the speed of propagation $\sfvMm$ of the domain
wall is given by
\bea
\sfvMm &=& \frac{\rhof-\betaM_m\rhoj}{\rhoj-\rhof}
\nonumber\\[2mm]
&=& \left( 1-\frac{\betaM_m}{\alpha} \right)
    \left( \frac{1-\betaM_m}{\nu}+\frac{1}{\alpha}-1 \right)^{-1}. 
\label{xvR1}
\eea
The second line results from some algebra in which (\ref{xJbetam})
and (\ref{xrhoJF}) are used;
the expression may be rewritten in several other ways.
When eliminating $\betaM_m$ from (\ref{xvR1}) and (\ref{xsfR})
we obtain the speed of propagation $\sfvMm$ of the reflected wave
as a function of $\sfRM_m$. Explicitly, 
\beq
\sfvMm = \frac{\alpha\nu\sfRM_m}{\alpha\sfRM_m+(1-\alpha)\nu}\,,
\label{xvR2}
\eeq
in which the coefficients are known functions of only the 
particle injection rate $a$.
This equation shows that we may retrieve the average domain wall
position from the reduced algorithm, in spite of the fact that 
at the microscopic level the waiting line has been eliminated from the
description.  
Equation (\ref{xvR2}) is valid only for $\alpha \geq \alphaM_m$.
When $\alpha$ decreases to $\alphaM_m$, we have that $\sfRMm\to 0$
and hence $\sfvMm\simeq [\alpha/(1-\alpha)]\sfRMm \to 0$.


\section{Monte Carlo results}
\label{sect_simulations}

Using the reduced algorithm of section \ref{sect_reduced}
we have carried out simulations
of the intersecting streets for values of $M$ up to $M=24$.
We determined the phase diagram
of the stationary state of the intersection square
as a function of the injection probability $\alpha$ 
and the linear size $M$.
In our figures we will present the reflection coefficients
$\sfRMm(\alpha)$, which by (\ref{relJmJf}) are directly equivalent to
the currents $\JM_m(\alpha)$.
Each simulation was started at $t=0$ from the initial random free flow 
configuration described in section \ref{sect_initialB}.
Relaxation to a stationary state appeared to be very rapid.

It is important to stress that, since we are in the limit $L=\infty$,
the simulation is {\it free of finite size effects.}
The remaining statistical errors in the simulation results
are entirely due to the finiteness of the simulation time. 
\vspace{3mm}


\begin{figure}
\begin{center}
\scalebox{.45}
{\includegraphics{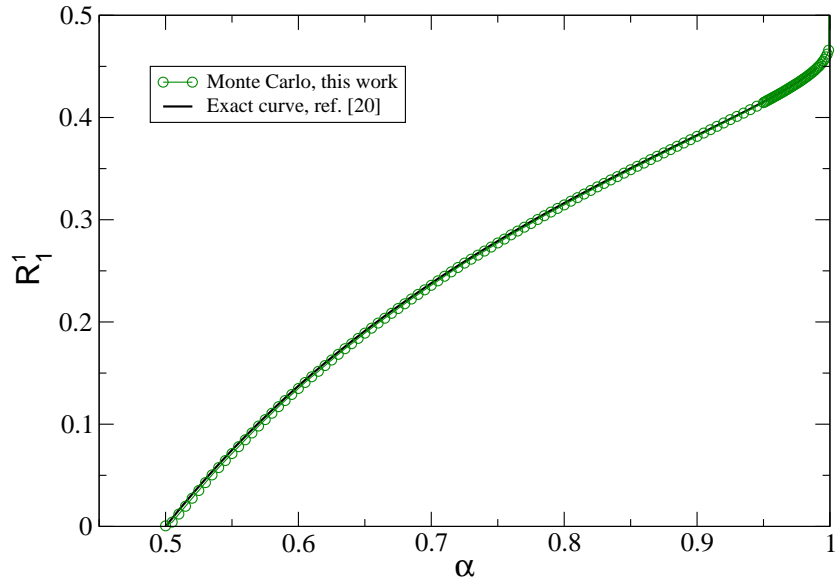}}
\end{center}
\caption{\small Reflection coefficient $\sfR^1_1(\alpha)$ for $M=1$.
The Monte Carlo algorithm of this work, which simulates an infinite
system, agrees very well
with the known analytic result \cite{ARCH11c}.}
\label{fig_RM1}
\end{figure}

\subsection{Phase diagram for $M=1$ and $M=10$}
\label{sect_phasediagrams}

For $M=1$ the intersection square consists of a single site
and there is a single critical jamming point $\alpha^1_1=\tfrac{1}{2}$
as predicted in earlier work \cite{ARCH11c}.
The $M=1$ simulation involves
less than a dozen variables, in spite
of the fact that we simulate an infinite system.
Figure \ref{fig_RM1} shows the simulation results 
for the reflection coefficient $\sfR^1_1(\alpha)$.
Each data point results from an average over $1.1\times 10^6$ time
steps and over the $x$ and $y$ direction.
One observes that $\sfR^1_1$ is nonzero only for $\alpha>\tfrac{1}{2}$
and that it tends to $\tfrac{1}{2}$
with infinite slope when $\alpha\to 1$.
A different representation of figure \ref{fig_RM1} was obtained analytically%
\footnote{The $M=1$ system actually studied in \cite{ARCH11c}
was more general: it had four control parameters 
(two entrance and two exit rates) which could break the symmetry between
the two crossing lanes; in the present work the two streets are
symmetric.}
by Appert-Rolland {\it et al.} \cite{ARCH11c},
whose results $J^{\rm free}=a/(1+a)$
and $J_1^1=\nu/(2\nu+1)$ imply that
$\sfR^1_1= (\nu-\nu/a+1)/(2\nu+1)$. This theoretical curve 
is also shown in the figure;
its full agreement with the Monte Carlo data confirms the correctness
of the `reduced' algorithm of this paper.
\vspace{3mm}


\begin{figure}
\begin{center}
\scalebox{.55}
{\includegraphics{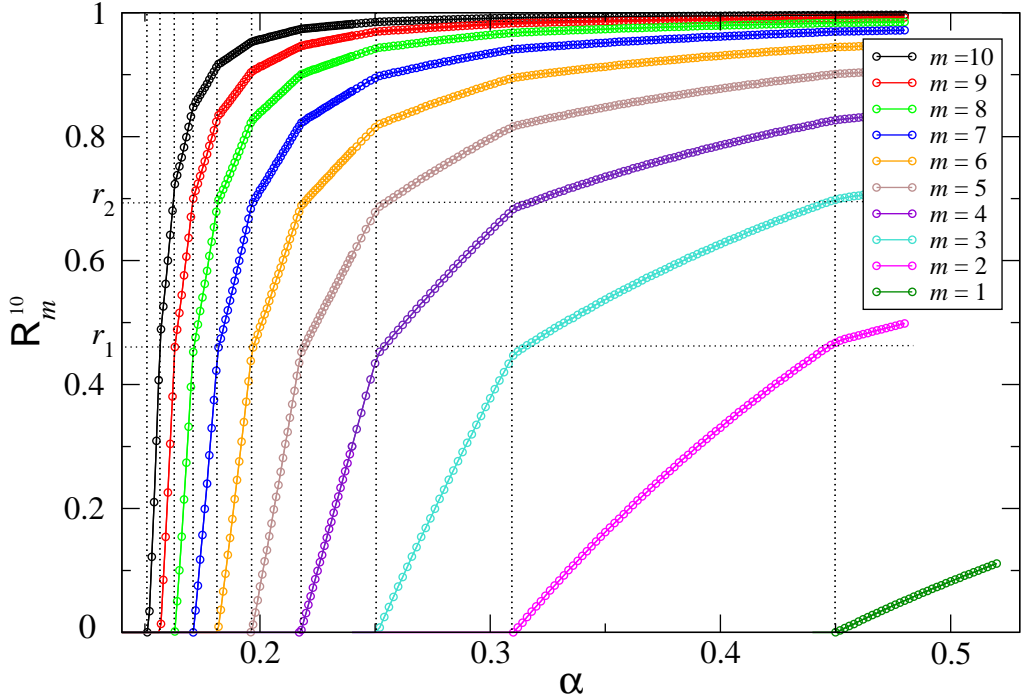}}
\end{center}
\caption{\small Reflection coefficient $\sfR^{10}_m(\alpha)$ for $M=10$
and $m=1,2,\ldots,10$. The lines connecting the points are meant to
guide the eye. The thin dotted vertical lines correspond to
the critical points on the $\alpha$ axis; in increasing order these are 
$\alphaten_{10}, \alphaten_9,\ldots,\alphaten_1$.
The thin dotted horizontal lines are the approximate levels at which the
curves have their first and second discontinuity of slope. 
}
\label{fig_RM10}
\end{figure}

Figure \ref{fig_RM10} is for $M=10$ and presents the 
reflection coefficients $\sfR^{10}_m(\alpha)$ for 
$m=1,2,\ldots,10$. Each data point results from an average over
$1.1\times 10^7$ time steps and the statistical error is less than the
symbol size.
There are ten critical values
$\alphaten_{10}<\alphaten_9<\ldots<\alphaten_1$.
For $\alphaten_{m+1}<\alpha<\alphaten_m$  the outer lanes $1,2,\ldots,m$
are in a free flow phase and the inner ones
$m+1,\ldots,M-1,M$ have jammed flow.
For each $m$ the data shown are averages on the two lanes $(m,x)$
and $(m,y)$. We verified that no spontaneous symmetry breaking occurs 
between the $x$ and the $y$ direction.


\subsection{Phase diagram for general M}
\label{sect_generalM}


\begin{figure}
\begin{center}
\scalebox{.45}
{\includegraphics{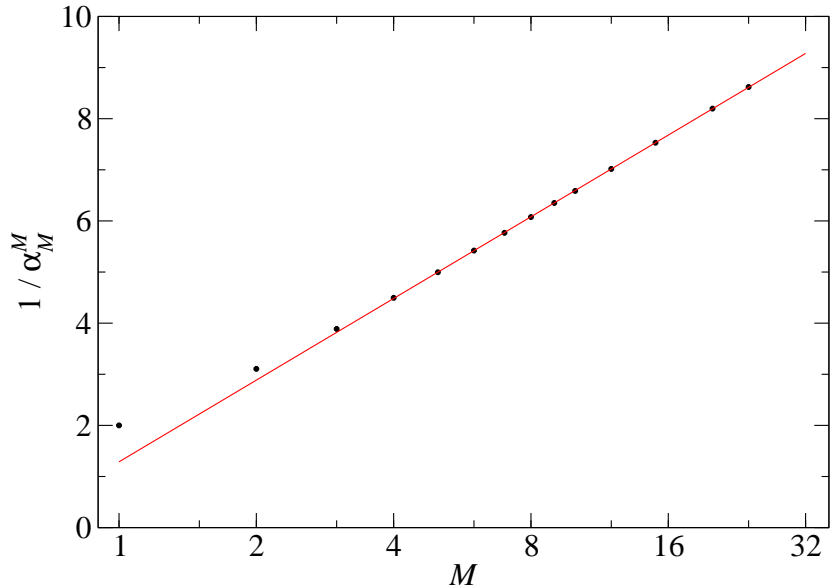}}
\end{center}
\caption{\small Inverse of the principal critical point $\alphaM_M$ 
as a function of the linear size $M$ of the intersection square
(notice the logarithmic scale of $M$).
The errors in the data points are smaller than the symbol sizes.
The straight line is the best asymptotic linear fit.}
\label{fig_alphaMM}
\end{figure}

\begin{figure}
\begin{center}
\scalebox{.45}
{\includegraphics{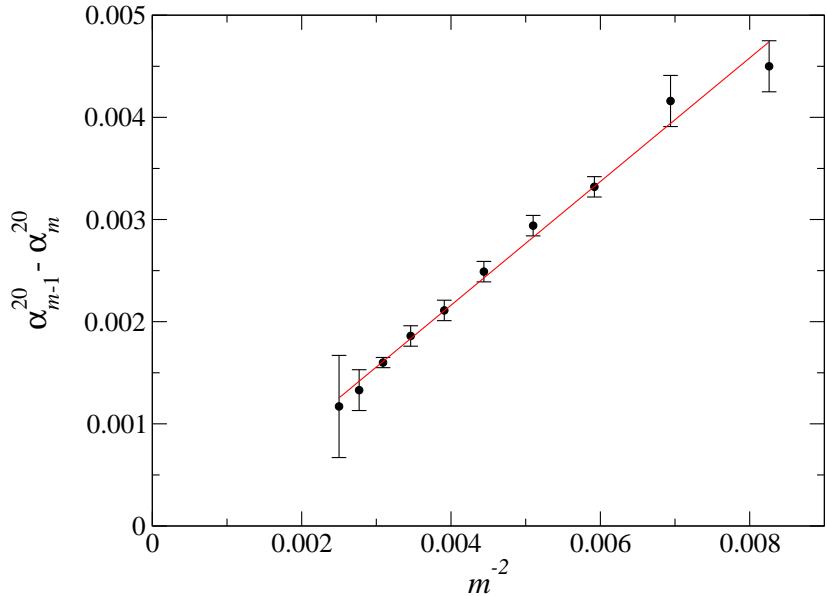}}
\end{center}
\caption{\small The intervals $\alphaM_{m-1}-\alphaM_{m}$ 
as a function of $m^{-2}$ for $M=20$. The straight line is the
best linear fit.}
\label{fig_alphadiff}
\end{figure}

We performed simulations for $M=1,2,\ldots,10,12,15,20,24$.
Our findings for the other $M$ values are qualitatively similar to those
obtained in figure \ref{fig_RM10} for $M=10$.
Again, there is no spontaneous symmetry breaking between the two
perpendicular directions.
As $\alpha$ increases, the system first reaches its principal critical
point $\alphaM_M$ at which the two innermost lanes (with $m=M$)
get jammed. Then there is an initially fast
succession of critical points 
$\alphaM_M<\alphaM_{M-1}<\alphaM_{M-2}<\ldots$
at each of which a further pair of lanes, one in each street, gets jammed.
The spacing between the critical points becomes gradually larger and
at the last critical point, $\alpha=\alphaM_1$, the outermost pair of
lanes (with $m=1$) gets jammed.

The principal critical point $\alphaM_M$ decreases with $M$
but its exact law is difficult to ascertain. It is very well
approximated by
\beq
\alphaM_M \simeq \frac{1}{A+B\log M}\,, \qquad \mbox{$M \gsim 4$},
\label{xalphaMM}
\eeq 
as shown in figure \ref{fig_alphaMM}, where $A=1.287$ and $B=2.306$.
The high precision of these results 
is due to the fact that even if the street {\it widths\,}
$M$ are finite, the simulations correspond
to infinite street {\it length,} $L=\infty$, and hence there are no
finite length effects.
The question of the large $M$ limit of the critical point was
asked also in the context of the BHL model
\cite{Bihametal92,Dingetal11}. 
All these authors have declared 
to be unable to state whether or not
this point goes to zero in the limit $M\to\infty$.
On the basis of our above results, albeit
for relatively small $M$ values, 
one might guess that (\ref{xalphaMM}) is the correct asymptotic law 
and hence that for the model of this work
the principal critical point does tend to zero with increasing $M$.
We refrain at present, however, from drawing this conclusion.
The reason is that
preliminary simulations for larger $M$ reveal complications
in the form of metastabilities associated with the transition possibly
turning first order. Since the infinite system
limit is a fundamental question in statistical mechanics,
we believe it is worthwhile, and necessary, to spend further 
efforts on studying it carefully.
\vspace{3mm}

The intervals between two successive critical points
scale in fairly good approximation as
\beq
\alphaM_{m-1}-\alphaM_{m} \,\simeq\, C_M + \frac{D_M}{m^2}\,,
\label{xdeltaalphaMm}
\eeq
where $C_M$ is a slightly negative and $D_M$ a positive constant.
Equation (\ref{xdeltaalphaMm}) appears to hold for $m$ not too small
and up to $m=M$. 
For $M=20$ this is shown in figure \ref{fig_alphadiff};
the fit has $C_{20}=-0.000\,26$ and $D_{20}=0.605$.

For $\alpha>\alphaM_m$ the curve $\sfRMm(\alpha)$ is composed of
$m$ segments that join with a discontinuity of slope
at the higher critical values $\alphaM_{m-1},\alphaM_{m-2},\ldots,\alphaM_1$.
For $M$ and $m$ not too small, these segments, although actually
curved, are very close to linear.
The slope of the first segment
of $\sfRM_m(\alpha)$, evaluated at $\alphaM_m$,
is approximately proportional to $m^2$.
Therefore, the larger $M$, the steeper the initial rise, 
and the finer should be the grid of points
on the $\alpha$ axis in order to obtain a resolution
of the successive jamming transitions in the inner lanes.
This renders the determination of $\sfRM_m(\alpha)$
by simulation increasingly harder as $M$ grows. 

The first discontinuity of slope
in each of the $\sfRM_m(\alpha)$ appears when  
$\sfRMm(\alpha)$ reaches a level $r_1$ that is, again in fairly good
approximation, $m$ independent. 
This had to be expected from the scaling 
observed above for the first segment of $\sfRM_m(\alpha)$,
whose horizontal and vertical extension are 
$\sim m^{-2}$ and $\sim m^2$), respectively.
For $M=10$ figure \ref{fig_RM10} shows that this level value is 
$r_1\approx 0.47$. 
Similarly, the second discontinuity appears at 
$\sfRMm(\alpha)\approx r_2\approx 0.69$, the third one 
(not explicitly indicated in the figure) at $r_3\approx 0.83$, and so on.
The values of these levels $r_i$ appear to depend only little on $M$, 
as anticipated in the notation.

As $M$ grows, the values $\alphaM_m$ with fixed $m$ seem to tend to
a limit value that may be called $\alpha^\infty_m$, but that we have
not tried to determine with any precision in this study.
Finally, for fixed $\alpha$ and $M$, the reflection
coefficient approaches unity, and hence the transmitted current
vanishes, roughly exponentially with $m$.
It is still a challenge to find an analytic explanation for all these
qualitative and quantitative observations, which were not expected 
{\it a priori.}  Whereas an exact solution seems beyond reach,
we think that an approximate theory may be possible.


\subsection{Snapshots}
\label{sect_snapshots}

\begin{figure}
\begin{center}
\scalebox{.70}
{\includegraphics{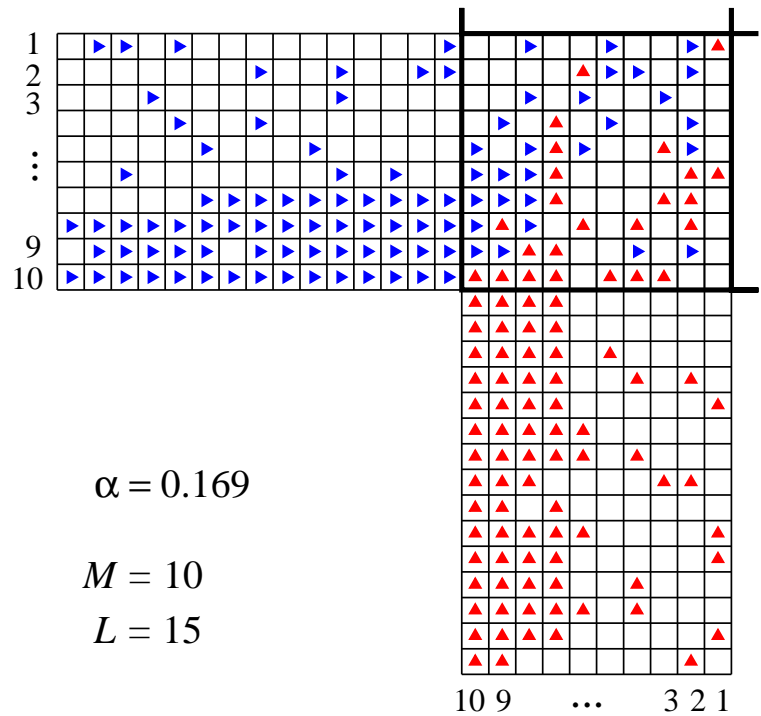}}
\end{center}
\caption{\small 
Snapshot of an intersection square
of linear size $M=10$ and for injection probability $\alpha=0.169$,
obtained by the `full' algorithm with $L=15$.
For this value of $\alpha$, lanes $1$ through $7$
are in the free flow and $8,9,10$ in the jammed flow phase.
}
\label{fig_snapshot10}
\end{figure}

Figure \ref{fig_snapshot10}
is a snapshot of a $10\times 10$ intersection square
simulated with an injection probability $\alpha=0.169$.
It is the only one of our simulations that was carried out with the
`full' algorithm. 
The length of the incoming streets is equal to $L=15$.
For this value of $\alpha$ figure \ref{fig_RM10} shows that the lanes
$m=1,2,\ldots,7$ are in the free flow phase
whereas those with $m=8,9,10$ are in the jammed flow phase. 
This is roughly what is visible in figure
\ref{fig_snapshot10}; nevertheless, considerable fluctuations occur
between independent snapshots, with 
lanes getting jammed and opening up intermittently.
These fluctuations are 
enhanced by the fact that for finite $L$ 
there are no sharply defined transition points, 
$L$ being automatically an upper limit on the length of the waiting lines.
\vspace{2mm}


\begin{figure}
\begin{center}
\scalebox{.70}
{\includegraphics{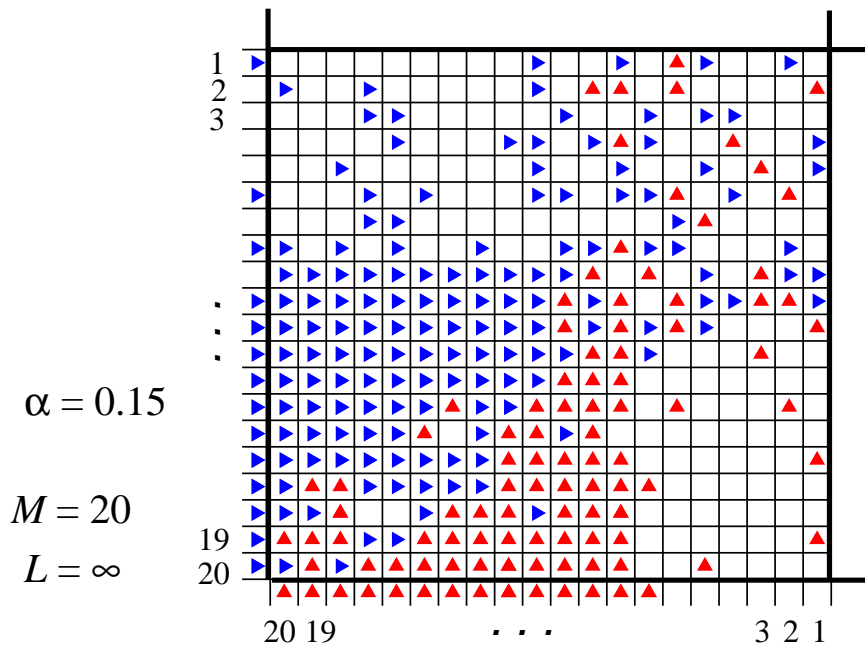}}
\end{center}
\caption{\small Snapshot of an intersection square
of linear size $M=20$ for injection probability $\alpha=0.15$,
obtained by the `reduced' algorithm ($L=\infty$).
The row and column of entrance sites are also shown.
For this value of $\alpha$ lanes $1$ through $10$
are in the free flow and $11$ through $20$ in the jammed flow phase.
}
\label{fig_snapshot20}
\end{figure}

Figure \ref{fig_snapshot20} is a snapshot of a $20\times 20$
intersection square together with its row and column of entrance sites.
The injection probability is $\alpha=0.15$.
For $M=20$ a phase diagram analogous to figure \ref{fig_RM10} shows
that $\alpha^{20}_{11}<0.15<\alpha^{20}_{10}$. As a consequence,
for $\alpha=0.15$ lanes $1$ through $10$ are in the free flow
phase and $11$ through $20$ are in the jammed flow phase.

Snapshots of this system taken at sufficiently long time intervals
show considerable variation, but all have in common the existence
of an approximately square (or sometimes rectangular) 
high density region in the lower left corner of the intersection square.
The shape of this corner region fluctuates, since the
jammed phase current still lets particles pass
relatively easily along theb outermost jammed lanes
(with indices $m=11,12,...$) of this region.
However, the jammed phase current
in the innermost lanes (those with indices $m=...,18,19,20$)
is very close to zero and the particles in those lanes 
are in a quasi-permanent frozen state. 
The net result is that there occurs a 
``freezing out'' of a set of inner lanes, which 
reduces the traffic problem on the $20\times 20$ intersection square 
to an effective one on a smaller square.
In figure \ref{fig_snapshot10}, this effective intersection
square would correspond to a square of size $10\times 10$
in the upper right corner.
This reduction is certainly not
exact but may well offer the starting point for a 
first theoretical approach.


\subsection{Further comments}
\label{sect_bhl}

Our algorithm for crossing streets of infinite length was reduced
to an algorithm on the finite-size intersection square 
with special memory boundary conditions.  
It is therefore natural that we make a comparison with existing
models on finite lattices.  A prominent one is
the BHL model due to Biham {\it et al.}
\cite{Bihametal92}. In this model 
two types of particles move unidirectionally on a torus of size
$M\times M$, one type horizontally and the other one vertically.
In the original version of this system the horizontally and vertically
moving particles are updated in parallel at the even and odd time
steps, respectively. 
A single phase transition was observed between a free flow phase
and a fully jammed (zero flow) phase.
The importance of the aspect ratio of
the lattice was stressed by D'Souza \cite{DSouza05}, who also showed
that there exists a third, high density, jammed phase.
In reference \cite{Bihametal92} and in the work 
that it has sparked 
\cite{Cuestaetal93,Moleraetal95,Nagatani93,Benyoussefetal03,DSouza05},
several variants of this BHL model were studied, many of them introducing
additional stochastic elements. 
Of particular interest in our context is a recent brief report
by Ding {\it et al.} \cite{Dingetal11}.
These authors considered the BHL model with the {\it standard\,} 
open boundary conditions that consist in filling
an empty site on the left or bottom boundary
with a fixed probability at each time step. 
Under such boundary conditions there is no place for waiting lines
and no sharply defined phase transition can be expected as long as $M$
is finite.
Obviously, the two types of boundary conditions correspond to distinct
driving parameters and have distinct sets of
applications: urban road traffic for the BHL model and intersecting
pedestrian traffic flows for ours.

The model of reference \cite{Dingetal11} also differs from ours
by its use of random sequential update; but although the update type is
important for the interpretation of the model, we consider that
difference as secondary.
\vspace{3mm}

The most striking similarity between the results of reference
\cite{Dingetal11} and ours
is that when the driving parameter increases, the $M\times M$ domain 
is gradually filled with a dense phase 
that occupies a square or rectangular lower left corner, 
in the way shown in our figure \ref{fig_snapshot20}.
We find, as in reference \cite{Dingetal11}, that
the complementary upper right corner,
which is in a free flow state, has a size roughly independent of $M$.

As for the differences,
our $M\times M$ domain is part of two infinite streets and the memory
boundary conditions keep track of the waiting lines in each of the $2M$
lanes. These boundary conditions create long-range
correlations in time, that in turn determine
sharply defined phase transition
points on the $\alpha$ axis for all $M=1,2,3,\ldots$.
As a consequence, in the model of this paper the progressive growth
of the dense corner occurs through a sequence of $M$ phase
transitions.
Our simulations concern relatively modest values of $M$
and are aimed at locating the transition points
with high precision. In addition  
we are able to assign to each of the $2M$ lanes 
a reflection coefficient which serves as a lane order parameter,
and a waiting line of which at each instant of time the average 
length is known.


\section{Conclusion}
\label{sect_conclusion}

We have introduced and studied a lattice model of
pedestrian traffic on two crossing one-way streets.
Each street is represented by a set of $M$ parallel TASEP lanes and
the only model parameter besides $M$
is the injection probability $\alpha$ of a pedestrian (`particle')
at minus infinity.
The dynamics is based on frozen shuffle update 
\cite{ARCH11a,ARCH11b,ARCH11c}: particles enter
stochastically but once in the system move deterministically.
From an algorithmic point of view we have found that in this model
the frozen shuffle update leads 
to accurate results even with a  modest simulation effort.
We consider this as an encouragement for simulating other
models, possibly very different ones, with the same update.

The intersecting streets that we considered
are infinite in both directions,
but one achievement of this work has been to
show that the dynamics may be reduced 
to a problem of interacting variables
on the sites of the finite $M\times M$ intersection square.
Appropriate boundary conditions were formulated in terms of `memory
variables' and we established the theoretical relation between
these new quantities and the outgoing current.

Our Monte Carlo work shows that as $\alpha$ increases,
the system undergoes a sequence of $M$ phase transitions
starting with the one at the principal critical point $\alphaM_M$.
Since we are able to perform the simulation directly on an infinite
system, there are no finite size effects; the uncertainty in the
critical points is due only to the finiteness of the simulation time.
At each transition a new pair of lanes, one in each street, passes from
a free flow to a jammed flow phase. We find that a reflection
coefficient for the current is an appropriate order parameter.
The accuracy of the algorithm has allowed us to 
establish certain surprising features, such as the discontinuities of slope
that occur when the reflection coefficient crosses a set of 
narrowly defined level values.
All these results should help trigger interest in building theoretical
approaches to this and similar systems.

The present study stays far from exploring, let alone answering, 
all questions that this specific model poses.
Quantities of interest not studied here include the particle density
in the intersection square, its space dependence, and its
fluctuations and correlations; 
the fluctuation of the waiting line lengths at and near the critical points;
and its relation to the fluctuation of the memory variables.
An unanswered question also concerns the behavior of the
principal critical point $\alphaM_M$ as $M\to\infty$.

We believe that this model, because of its simplicity, 
sets a standard scenario with respect to which others
may be discussed.
Clearly, ideas come readily to mind about how to 
modify or extend this model, 
for example by considering streets of different widths
$M_1$ and $M_2$, unequal injection probabilities 
$\alpha_x$ and $\alpha_y$, and so on;
or by opening the possibility for
pedestrians to move laterally or diagonally forward.
Sideways steps between the lanes are certainly expected to blur
the distinction between the jamming transitions in the individual
lanes that here occur separately.
It is hard to guess, however, what sort of a transition will then
result. 
Several effects of the present model may turn out to be robust, 
such as the predominance of the
flow through the outer lanes over those through  
the inner ones. In future work \cite{CARH12}
we will address a few of the many new issues raised here.


\section*{Acknowledgments}

The authors thank L. Santen for pointing out to them several important
references. They also thank
J. Cividini, J.-M. Caillol, and R.K.P. Zia for their interest 
shown during various stages of this work.


\end{document}